\documentclass[runningheads]{llncs}
\usepackage[T1]{fontenc}   
\usepackage{amsmath}
\usepackage{graphicx}
\newtheorem{fact}{Fact}
\newcommand{\junk}[1]{}

\begin{document}

\title{Improved Lower Bounds for Monotone $q$-Multilinear Boolean Circuits}

\author{Andrzej Lingas
  \inst{1}
  \and
  Mia Persson
  \inst{2}
}
\authorrunning{A. Lingas and M. Persson}
\institute{
Department of Computer Science, Lund University, 22100 Lund, Sweden. 
\email{Andrzej.Lingas@cs.lth.se}
\and
Department of Computer Science and Media Technology, Malm\"o University, 20506 Malm\"o, Sweden.
\email{miapersson@mau.se}}
\maketitle
\begin{abstract}
  A monotone Boolean circuit is composed of OR gates, AND gates and
  input gates corresponding to the input variables and the Boolean
  constants. It is $q$-multilinear if for each its output gate $o$ and
  for each prime implicant $s$ of the function computed at $o$, the
  arithmetic version of the circuit resulting from the replacement of
  OR and AND gates by addition and multiplication gates, respectively,
  computes a polynomial at $o$ which contains a monomial including the
  same variables as $s$ and each of the variables in $s$ has degree at
  most $q$ in the monomial.

  First, we study the complexity of computing semi-disjoint bilinear
  Boolean forms in terms of the size of monotone $q$-multilinear
  Boolean circuits.  In particular, we show that any monotone
  $1$-multilinear Boolean circuit computing a semi-disjoint Boolean
  form with $p$ prime implicants includes at least $p$ AND gates.  We
  also show that any monotone $q$-multilinear Boolean circuit
  computing a semi-disjoint Boolean form with $p$ prime implicants has
  $\Omega(\frac p {q^4})$ size.

  Next, we study the complexity of the monotone Boolean function
  $Isol_{k,n}$ that verifies if a $k$-dimensional Boolean matrix has at
  least one $1$ in each line (e.g., each row and column when $k=2$), in
  terms of monotone $q$-multilinear Boolean circuits. We show that
  that any $\Sigma_3$ monotone Boolean circuit for $Isol_{k,n}$ has an
  exponential in $n$ size or it is not $(k-1)$-multilinear.
\end{abstract}
\footnote{This  manuscript is a substantial extension and
  improvement of~\cite{L23}.}
\vfill
\newpage
\section{Introduction}
Prospects for deriving superlinear lower bounds on the size
of Boolean circuits for natural problems are very weak.
For this reason, already by the end of the 70s and the beginning of the 80s,
one started to study the complexity of monotone arithmetic
circuits and/or monotone Boolean circuits for natural multivariate arithmetic
polynomials and
natural Boolean functions, respectively.
The monotone arithmetic circuits are composed of addition gates, multiplication
gates and input gates for variables and non-negative real constants.
Similarly, monotone Boolean circuits are composed of OR gates, AND gates.
and the input gates for variables and Boolean constants.
In the case of monotone arithmetic circuits, one succeeded to show
even exponential lower bounds relatively easily \cite{JS82,S76} while in the case
of monotone Boolean circuits, the derivation of exponential
lower bounds for natural problems required more effort \cite{AB87,razb-clique}.

In this context,
the problem of computing the permanent of an $n\times n$ $0-1$
matrix equivalent to counting the number of perfect matchings
in a bipartite graph is very interesting.
Jerrum and Snir established
an exponential lower bound on the size of monotone arithmetic
circuits for this problem \cite{JS82} while the best known lower
bound on the size of a monotone Boolean circuit
for the Boolean variant of the permanent due to
Razborov~\cite{razb-perm} is only superpolynomial. In order to tackle
the gap, Ponnuswami  and Venkateswaran considered the
concept of monotone multilinear Boolean circuits  and showed an
exponential lower bound on the size of such circuits for the Boolean permanent \cite{KV04}.
They call a monotone Boolean circuit  multilinear if
for any its AND gate the two input functions have no variable in common.
\junk{To be more precise, the authors of
\cite{KV04} used a semantic version
of multilinearity by forbidding  minimal representations of the two input functions
to share a variable \cite{KV04}.
This means for example that if the first function is represented by
$x\vee x\wedge y$ and the second one by $y$ then the AND gate
represented by $(x\wedge y )\vee (x\wedge y \wedge y)$ is allowed in their multilinear Boolean circuit.}
Soon after, Krieger obtained exponential lower bounds
on the number of OR gates in monotone multilinear Boolean circuits for among other things
a clique function \cite{K07}. He used a much more restricted syntactic
version of multilinearity. In this version, the function computed at a gate is declared to be
dependent on a variable if there is a path from the input gate with the variable
to the gate in the circuit. On the other hand, his lower bounds include also
DeMorgan multilinear Boolean circuits. These circuits are a generalization
of monotone multilinear Boolean circuits, allowing for the restricted use of negation operation
that can be applied only to the input variables.

In a recent report \cite{L22}, Lingas used a simple argument to obtain
in a way a more general result than the
lower bounds of Ponnuswami and Venkateswaran or Krieger in \cite{KV04}
and \cite{K07}, respectively.  He has shown
that the known lower bounds on the size of monotone arithmetic
circuits for multivariate polynomials that are sums of monomials
consisting of the same number of distinct variables \cite{JS82,S76} yield
almost the analogous lower bounds on the size of monotone multilinear Boolean
circuits computing the functions represented by the corresponding
multivariate Boolean polynomials. His result has been slightly improved
by Jukna in \cite{SJ22} who showed that exactly analogous lower bounds 
can be obtained by using the lower envelope argument from \cite{JS82}.

One should also mention that Raz and Widgerson showed that monotone
Boolean circuits for the Boolean permanent require linear depth \cite{raz-wigderson92} and Raz
proved that multilinear Boolean formulas for this problem have superpolynomial size \cite{Raz03}.

The concept of circuit multilinearity is also natural for circuits
over other semi-rings beside the Boolean one $(\{0,\ 1\}\, \vee,
\wedge)$ such as the arithmetic one $(R_+ , +, \times)$ or the
tropical one $(R_+, \min, +)$, where $R_+$ stands for the set of
nonnegative real numbers \cite{SJ22}.  In particular, Jukna observed
that the classical dynamic programming algorithms for shortest paths
and traveling salesperson problems can be expressed as multilinear
circuits over the tropical semi-ring \cite{SJ22}.

In this paper, we consider a generalization of monotone multilinear
Boolean circuits to include monotone $q$-multilinear Boolean
circuits. A monotone Boolean circuit is $q$-multilinear if for each
its output gate $o$ and for each prime implicant $s$ of the function
computed at $o$, the arithmetic version of the circuit resulting from
the replacement of OR and AND gates by addition and multiplication gates,
respectively, computes a polynomial at $o$ which contains a monomial
including the same variables as $s$ and each of the variables in $s$ has
degree at most $q$ in the monomial.

The central question is how restrictive is the requirement of
$q$-multilinearity in monotone Boolean circuits.  In particular,
whether or not there are substantial gaps between the sizes of
monotone $q$-multilinear Boolean circuits and the sizes of the
monotone $(q+1)$-multilinear Boolean circuits computing the same
Boolean functions.

Monotone $q$-multilinear circuits correspond to
the so called monotone read-$q$ Boolean circuits introduced by Jukna in
\cite{SJ22}. He showed in \cite{SJ22} that monotone $1$-multilinear
(i.e., read-$1$) circuits coincide with monotone multilinear
Boolean circuits. He also proved that the function $Isol_n$ that verifies
if an input $n\times n$ Boolean matrix has at least one $1$
in each of its rows and columns admits a monotone
$2$-multilinear Boolean circuit of linear size but any
monotone $1$-multilinear Boolean circuit computing $Isol_n$
has an exponential in $n$ size.

First, we study the complexity of computing semi-disjoint bilinear
Boolean forms.
Boolean matrix product and Boolean vector convolution  are
the best known examples of semi-disjoint bilinear Boolean forms.
A central challenge here is to show that any monotone
  Boolean circuit computing a semi-disjoint bilinear Boolean form with
  $p$ prime implicants  has $\Omega(p)$ size. We address this
  challenge in terms of the size of monotone $q$-multilinear Boolean
  circuits.  In particular, we show that any monotone $1$-multilinear
  Boolean circuit computing a semi-disjoint Boolean form with $p$
  prime implicants includes at least $p$ AND gates.  We also show that
  any monotone $q$-multilinear Boolean circuit computing a
  semi-disjoint Boolean form with $p$ prime implicants has
  $\Omega(\frac p {q^4})$ size.  The latter bound applied to the
  $n$-dimensional Boolean convolution subsumes the best known lower
  bound of $\Omega (n^2/\log ^6 n)$ on the size of monotone Boolean
  circuits for this problem due to Grinchuk and Sergeev \cite{GS11} as
  long as $q=o(\log^{3/2} n).$

  Next, we address the intriguing question about the power of
  idempotency. A challenge here is to show that for every natural $q,$
  there is a Boolean function that admits a monotone
  $(q+1)$-multilinear Boolean circuit of polynomial size but any
  monotone $q$-multilinear Boolean circuits that computes this
  function has a superpolynomial size. As a candidate function, we
  choose the monotone Boolean function $Isol_{k,n}$ that is a
  generalization of the function $Isol_n$ considered in \cite{SJ22}.
  The generalized function
  verifies if a $k$-dimensional Boolean matrix has at least one $1$ in
  each line (e.g., each row and column when $k=2$).  We show that that
  any $\Sigma_3$ monotone Boolean circuit for $Isol_{k,n}$ has an
  exponential in $n$ size or it is not $(k-1)$-multilinear.  The
  latter result subsumes a corresponding result relying on
  additional assumptions, given in Theorem 6 in
  \cite{L23}.
  
\section{Monotone Boolean circuits and functions}
A {\em monotone Boolean circuit} is a finite directed acyclic graph with
the following properties.

\begin{enumerate}
\item The indegree of each vertex (termed gate) is either $0$ or $2.$
\item The source vertices (i.e., vertices with indegree $0$ called
input gates) are
labeled by variables or  the Boolean constants $0,\ 1.$
\item The vertices of indegree $2$ are labeled by elements
of the set $\{OR, AND\}$ and termed OR gates and AND gates,
respectively.
\item A distinguished set of gates forms the set of output gates of the circuit.
\end{enumerate}


For convenience, we denote also by $g$ the function computed at a gate $g$
of a monotone Boolean circuit.
The {\em size} of  a monotone Boolean circuit is
the total number of its non-input gates. 

A monotone Boolean circuit is {\em multilinear} if for any AND gate
the two input Boolean functions have no variable in common.  The {\em
  conjunction depth} of a monotone Boolean circuit is the maximum
number of AND gates on a path from an input gate to an output gate. 
The {\em alternation depth} of a monotone Boolean circuit is
the maximum number of blocks of consecutive OR gates and blocks of
consecutive AND gates on a
path from an input gate to an output gate.  A $\Sigma_d$ -circuit
(respectively, $\Pi_d$-circuit) is a circuit with the alternation depth
not exceeding $d$ such that the output gates are OR gates (AND gates,
respectively).

With each gate $g$ of a monotone Boolean circuit,
 we associate a set $T(g)$ of
terms in a natural way. Thus, with each input gate,
we associate the singleton set consisting of
the corresponding variable or constant. Next,
with an OR gate, we associate the union of
the sets associated with its direct predecessors.
Finally, with an AND gate $g,$ we associate the
set of concatenations $t_1t_2$ of all pairs of terms $t_1,$ 
$t_2,$ where $t_i\in T(g_i)$ and $g_i$ stands
for the $i$-th direct predecessor
of $g$, for $i=1,2.$
The function computed at the gate $g$ is 
the disjunction of the functions (called monoms)
represented by the terms in $T(g).$
The monom $con(t)$ represented by a term $t$ is obtained
by replacing concatenations in $t$ with conjunctions,
respectively.
A term in $T(g)$ is a {\em zero-term} if it contains
the Boolean constant $0$. Clearly, a zero-term
represents the Boolean constant $0.$
By the definition of $T(g)$ and induction on
the structure of the monotone  Boolean circuit,
$g=\bigvee_{t\in T(g)}con(t)$ holds.
For a term $t\in T(g),$ the set of variables occurring in $t$
is denoted by $Var(t)$.

A Boolean {\em form} is a finite set of Boolean 0-1 functions.
An {\em implicant} of a Boolean form $F$ 
is a conjunction
of some variables
and/or Boolean constants (monom)
such that there 
is a function belonging to $F$
which is true whenever the conjunction is true.
If the conjunction includes the Boolean $0$   then it is
a {\em trivial implicant} of $F.$
A non-trivial implicant of $F$ that is minimal with
respect to included variables  is a {\em prime implicant}
of $F.$ The set of prime implicants of $F$ is denoted
by $PI(F).$

A (monotone) {\em Boolean polynomial} is a disjunction of 
  monoms, where each monom is a conjunction of some variables and
Boolean constants.  It is a {\em minimal} Boolean polynomial
representing a given Boolean function if after the removal of any
variable or constant occurrences, it does not represent this function.

A set $F$ of monotone Boolean functions
is a {\em semi-disjoint bilinear form}
if it is defined on the set of variables
$X\cup Y$ and the following properties hold.

\begin{enumerate}
\item For each minimal Boolean polynomial
  representing a Boolean function
  in $F$ and each variable
$z\in X\cup Y,$ there is at most 
one monom
of the polynomial containing $z.$ 
\item 
  Each monom of a minimal Boolean polynomial
  representing a Boolean function in $F$ consists of exactly
one variable in $X$ and one variable in $Y.$
\item The sets of monoms of minimal Boolean polynomials
  representing different Boolean functions in $F$
are pairwise disjoint.
\end{enumerate}

Boolean matrix product and Boolean vector convolution  are
the best known examples of semi-disjoint bilinear Boolean forms.

\section{Monotone $q$-multilinear Boolean circuits}
Recall that a monotone Boolean circuit is multilinear
if for any of its AND gates the two input functions
do not share a variable.

The following lemma provides a characterization
of the terms produced at the gates of a monotone  multilinear circuit
which lays  ground to the generalization of the multilinearity
to include the $q$-multilinearity. To specify the lemma, we need to introduce
the following additional notation.

Let $g$ stand for a gate of a monotone multilinear circuit.  For two
terms $t,\ t'\in T(g),$ the relationship $t'\le t$ holds if and only
if for each variable $x$, the number of occurrences of $x$ in $t'$
does not exceed that in $t.$ A {\em variable repetition} takes place
in $t$ if there is a variable which occurs at least two times in
$t.$
  
\begin{lemma}\label{lem: comp}(companion lemma)
  Let $g$ be a gate of a monotone multilinear Boolean circuit
  without the Boolean constants, and let $t\in T(g).$
  There is $t'\in T(g)$ without
  variable repetitions such that $t'\le t.$
  \end{lemma}
    \begin{proof}
      The proof is by induction on the structure of the circuit in a bottom-up manner.
      If $g$ is an input gate corresponding to a variable 
      then $t'=t.$
      If $g$ is an OR gate then the lemma for the gate immediately follows
      from the induction hypothesis. Suppose that $g$ is an AND gate with
      two direct gate predecessors $g_1$ and $g_2.$ Consider $t=t_1t_2\in T(g),$
      where $t_i\in T(g_i)$ for $i=1,\ 2.$ By the induction hypothesis,
      there are (non-zero) terms $t_i'\in T(g_i)$ without variable repetitions such that
      $t_i'\le t_i$ for $i=1,\ 2.$ Let $t'=t_1't_2'.$ It follows that $t'\le t.$
      If $t'$ has a variable repetition then there exist a variable $x$
      and $j\in \{ 1,2\}$ such that $t_j'$ has an occurrence of the variable
      but the function $g_j$ does not depend on $x.$
      Hence, there must exist a term $t_j''\in T(g_j)$ without an occurrence of  $x$ 
      such that the monom represented by $t_j''$ is implied
      by $t_j',$ i.e., $Var(t_j'')\subset Var(t_j').$
      We may assume without loss of generality that $t_j''$
      does not contain variable repetitions since otherwise we can replace
      it with a smaller term with respect to $\le $ without variable repetitions by the induction
      hypothesis. We may also assume
      without loss of generality that $j=1.$ Hence, $t_1''t'_2\le t$ and if $t_1''t_2'$ is
      free from variable repetitions we are done. Otherwise, we repeat
      the procedure eliminating next variable on
      one of the sides. Because the number of variables is finite the process
      must eventually result in a term satisfying the lemma.
      \qed
    \end{proof}

\junk{
Fix a positive integer $q.$ Roughly, a sufficient condition on a Boolean
monotone circuit to call it $q$-multilinear mentioned in the
introduction is that in the formal Boolean polynomials at the output
gates of the circuit, no variable has multiplicity larger than $q.$
More formally, in terms of our notation it can be rephrased
as follows:

In each term in $T(g),$ where $g$ is an output gate, each variable
has at most
$q$ occurrences. We shall call a monotone Boolean circuit
satisfying this condition {\em strictly $q$-multilinear}.

Our definition of a monotone $q$-multilinear Boolean circuit
imposes a weaker condition and it corresponds to that of
a monotone read-$k$ Boolean circuit from \cite{SJ22}.
}
A monotone Boolean circuit computing a monotone Boolean form $F$ is said to be
{\em $q$-multilinear} if for each prime implicant $p$ of each function
$f\in F,$ there is a term $t\in T(o)$ representing $p,$ where $o$ is
the output gate of the circuit computing $f$, such that no variable
occurs more than $q$ times in $t.$ Moreover, if for each gate $g$ of the circuit,
no term in $T(g)$ contains more than $q$ occurrences of a single variable
then the circuit is called {\em strictly q-multilinear}.

Our definition of a monotone $q$-multilinear Boolean circuit
corresponds to that of a monotone read-$q$ Boolean circuit from
\cite{SJ22}.

By the companion lemma
or Lemma 4
in \cite{SJ22}, any monotone multilinear Boolean circuit is $1$-multilinear.
For the reverse implication (in terms of monotone read-$1$ Boolean
circuits), see also Lemma 4 in \cite{SJ22}. Among other things
because of the aforementioned equivalence, we believe that the name
``$q$-multilinear'' is more natural than the name ``read-$q$'' used in \cite{SJ22}.

When a monotone $q$-multilinear Boolean circuit computes a Boolean form
with prime implicants  of length $k$ then by the following theorem it
can be transformed into a monotone
strictly $(k(q-1)+1)$-multilinear Boolean circuit.
\begin{theorem}
 Let $C$ be a monotone $q$-multilinear Boolean circuit
 computing a Boolean form $F.$ Let $s$ be the size
 of $C$ and let $m$ be the maximum
 number of variables in a prime implicant of $F.$
 The circuit $C$ can be transformed into a monotone
 $q$-multilinear Boolean circuit $C'$ computing
 $F$  such that the size of $C'$ is $O(m^2q^2s)$
 and no term produced by $C'$ has more than $mq$
 variable occurrences. Consequently, the circuit
 $C'$ is strictly $(mq-m'+1)$-multilinear, where
 $m'$ is the minimum number of variables in a
 prime implicant of $F.$
\end{theorem}
\begin{proof}
  We may assume w.l.o.g. that $C$ does not use Boolean constants.
  The idea of the construction of $C'$ on the basis of $C$
  is as follows. For each gate $g$ of $C$, there are at most
  $mq$ corresponding gates $g_1,..,g_{mq}$ in $C'$.
  For $1\le \ell \le mq,$ the
  set of terms in $T(g_{\ell})$ is supposed to consist of the
  (non-zero) terms in $T(g)$ having exactly $\ell$ variable occurrences.

  The construction of $C'$ is straightforward.  For an input gate $g$
  corresponding to a variable $x$, only $g_1=x$ is defined.  Consider
  a gate $g$ of $C$ with direct predecessor gates $g',\ g''$ and $1\le
  \ell \le mq.$ If $g$ is an OR gate then $g_{\ell}=g'_{\ell} \vee
  g''_{\ell}$, provided that both $g'_{\ell}$ and $g''_{\ell}$ are
  defined. If only one of the latter gates is defined then it is
  substituted for $g_{\ell},$ if none then $g_{\ell}$ is undefined.
  Thus, there are at most $mq$ OR gates in $C'$
  corresponding to an OR gate in $C.$ If $g$ is an AND gate then
  $g_{\ell}=\bigvee_{j=1}^{\ell -1}g'_j\wedge g''_{\ell -j}$, where
  the conjunction $g'_j\wedge g''_{\ell -j}$ takes place if both
  $g'_j$ and $g''_{\ell -j}$ are defined for $j=1,...,\ell -1.$ If no
  conjunction takes place in the disjunction then $g_{\ell}$ is undefined.
Thus, in the case of the AND gate, a partial convolution of
    $(g'_1,...,g'_{mq})$ and $(g''_1,...,g''_{mq})$ needs to be computed.
It requires $O((mq)^2)$ AND and OR gates.
For each output gate $g$ of $C,$ the corresponding output gate of $C'$
    computes the disjunction of the defined gates $g_{\ell},$
    $1\le \ell \le mq.$

    It follows by the construction of $C'$ that it has size
    $O(m^2q^2s),$ and produces terms having at most $mq$ occurrences
    of variables. It remains to show that $C'$ computes the form $F$,
    it is $q$-multilinear, and strictly $(mq-m'+1)$-multilinear.

    Consider an output gate $g'$ of $C'$ corresponding to an output gate $g$ of
    $C$. By the definition of $g',$ each term in $T(g')$ represents an implicant
    of the function computed at $g.$ By the $q$-multilinearity of $C$
    and the definition of $g',$ for each prime implicant of the function
    computed at $g,$  $T(g)$ and consequently $T(g')$ contain
    a term representing this prime implicant such that no variable in
    the term occurs more than $q$ times. It follows that $g'$ computes
    the same Boolean function as $g$, and that $C'$ is also $q$-multilinear.

    Finally, since each output term $t$ produced by $C'$
    contains at most $mq$ variable occurrences and represents an implicant of $F,$
    no variable in $t$ can be repeated more than $mq-m'$
    times. Thus, $C'$ is strictly $(mq-m'+1)$-multilinear.
    \qed
\end{proof}

Recall that a monotone Boolean circuit is of conjunction depth
$d$ if the maximum number of AND gates on any path from
an input gate to an output gate in the circuit is $d.$
A bounded conjunction depth yields a rather weak upper bound on
the $q$-multilinearity of a monotone Boolean circuit.
  \begin{theorem}
    Let $F$ be a monotone Boolean form, and let $k$ be the minimum
   number of variables forming a prime implicant of $F.$
   A monotone Boolean circuit of conjunction  depth $d$
   computing $F$ is strictly $(2^d-k+1)$-multilinear.
  \end{theorem}

  \begin{proof}
    An AND gate can at most double the maximum length of the terms (i.e., the
    number of variable occurrences in the terms) produced by its
    direct predecessors. Hence, the output terms of a monotone Boolean
    circuit of conjunction depth $d$ have length not exceeding $2^d.$
    Consider a variable $x$ occurring in an output term of a monotone
    Boolean circuit of conjunction depth $d$ computing $F.$ As the term
    represents an implicant of $F,$ it has to contain at least $k-1$
    other variables. Hence, the maximum number of occurrences of $x$
    in the term is $2^d-k+1.$
    \qed
  \end{proof}
  
The reverse relationship is much stronger.
  
  \begin{theorem}\label{fact: depth}
    Let $C$ be an optimal monotone $q$-multilinear Boolean circuit
    without the Boolean constants computing a
  monotone Boolean form $F$ whose prime implicants are formed
  by at most $k$ variables.
  The circuit has conjunction depth not exceeding $kq-1.$
  \end{theorem}
\begin{proof}
Consider terms at the output gates of $C$ representing prime
  implicants of $F.$ We know that for each prime implicant $p$ of $F$
  there is a term $t_p$ representing $p,$ where each variable
  occurs at most $q$ times. Consider the sub-dag $C_p$ of the circuit
  generating the term $t_p.$ Note that $C_p$ includes the input
  gates corresponding to the at most $k$ variables in $t_p$ and for any OR
  gate included in $C_p$ exactly one of the direct predecessors gates
  in $C$ is included (such a sub-dag is termed {\em parse graph} in
  \cite{SV94}).  Let $P$ be a path from an input gate
  labeled by a variable $x$ to the output gate
  in the sub-dag having the maximum number of AND gates.  Note that at
  each AND gate $h$ on the path $P$, a subterm of $t_p$ including $x$
  and belonging to $T(h)$ has to be larger at least by one variable
  occurrence than that belonging to the direct predecessor of $h$ on
  the path. We conclude that there are at most $kq-1$ AND gates on the
  path $P.$

Form the sub-dag $C'$ of $C$ that is the union of the sub-dags
$C_p,\ p\in PI(F).$ Note that some OR gates in $C'$ may have only one
direct predecessor, we replace the missing one with the Boolean $0.$
Let $g'$ be the output gate of $C'$ corresponding to the output gate
$g$ of $C.$ By the definition, $T(g')$ includes terms representing all
prime implicants of $F$ represented in $T(g).$ Consider a non-zero
(i.e., not including $0$) term $t\in T(g')$ that does not represent a
prime implicant of $F.$ Consider the sub-dag (parse-graph) $C'_t$ of
$C'$ that generates exactly the term $ t.$ It also generates the term
$t$ in the original circuit $C.$ We conclude that $t$ is an implicant
of $F$ and consequently that $C'$ computes $F.$
By the definition,
$C'$ has conjunction depth bounded by $kq-1,$ size not exceeding that of $C$,
and it is also $q$-multilinear.
Since the Boolean constants can be eliminated from $C'$
decreasing its size, we conclude that $C'$ has the same size as $C$
by the optimality of $C$ and consequently 
that $C=C'$ by the construction of $C'.$
\qed
\end{proof}

\section{Lower bounds for $q$-multilinear Boolean circuits}

\subsection{Lower bounds for semi-disjoint bilinear forms}

For the Boolean matrix product of two $n\times n$ Boolean matrices,
there are known tight $\Theta(n^3)$ bounds on the number of AND and OR
gates in monotone Boolean circuits computing the product \cite{MG76,Pat75,P75}.
They yield analogous bounds on the number of AND and OR gates in
monotone $1$-multilinear Boolean circuits for the product.
In case of Boolean convolution of two $n$-dimensional Boolean
vectors, the tight lower bounds on the number of additions and multiplications in monotone
arithmetic circuits computing the arithmetic convolution
of two $n$-dimensional vectors \cite{JS82,S76} translate
to the analogous bounds on the number of OR gates and AND gates in monotone
multilinear Boolean circuits computing the Boolean vector convolution,
by the general equivalence established in \cite{SJ22}.

We shall derive a general lower bound on the number of AND gates in a
monotone $q$-multilinear Boolean circuit computing a semi-disjoint
bilinear Boolean form with $p$ prime implicants.
However, first we shall derive a tight lower
bound in the case $q=1,$ showing that $p$ AND gates are needed then.
For this purpose, we need the following lemma.

\begin{lemma}\label{lem: 2}
Let $C$ be a Boolean circuit 
computing a semi-disjoint bilinear form $F$
on the variables $x_0,...,x_{n-1}$ and $y_0,...,y_{n-1}.$ 
Let $h$ be an AND gate computing the conjunction
of the functions computed at the gates $h_1$and $h_2,$
respectively. Suppose that 
the single variables
$u_1,u_2$ belong to the set of
prime implicants of the function computed at
$h_1$ while the single variable $w_1$ belongs
to the set of prime implicants of the function
computed at $h_2$ so that $u_1w_1$ is a
prime implicant of $F$.
Let $o$ be an output gate computing the function in $F$
for which $u_1w_1$ is a prime implicant.
If a term in $T(h)$ representing $u_1w_1$ is a part
of a term in $T(o)$ representing $u_1w_1$ in which
each variable occurs at most once then
$C$ cannot be $1$-multilinear.
\end{lemma}
\begin{proof}
  For $i=1,2,$ let $t(u_iw_1)$ be the term in $T(h)$ representing $u_iw_1$.
  We may assume w.l.o.g. that the term in $T(o)$ representing
  $u_1w_1$ has the form $t_1t(u_1w_1)t_2$ and that no variable occurs more
  than once in it.
  It follows that the only variables that $t_1t_2$ could include
  are $u_1$  and/or $w_1.$ But then at least one variable would
  occur two times in $t_1t(u_1w_1)t_2$, contradicting our assumptions. If $t_1,\ t_2$ are empty words or equivalent
  to the Boolean $1$ then $t_1t(u_2w_1)t_2$ does not represent
  an implicant of the function computed at $o$ by the definition
  of a semi-disjoint bilinear Boolean form (first condition). We obtain
  a contradiction.
  \qed
\end{proof}

Our first main result is as follows.

\begin{theorem}
  A monotone $1$-multilinear Boolean circuit computing
  a semi-disjoint bilinear Boolean form with $p$ prime implicants
  has at least $p$ AND gates.
\end{theorem}
\begin{proof}
 Let $C$ be a monotone $1$-multilinear Boolean circuit computing
 the semi-disjoint bilinear form $F$ with $p$ prime implicants.
 To each prime implicant $s$ of $F$, we can assign a term $a(s)$
 in $T(o),$ where $o$ is the output gate of $C$ that computes
 the function whose set of prime implicants includes  $s$, such that $a(s)$
 represents $s$ and no variable occurs twice or more times
 in $a(s).$
 Next, for an AND gate $h$ of $C,$
 let $S_h$ denote the set
of prime implicants $s$ of $F$ such that:
\begin{enumerate}
\item
 $s$ is a prime implicant
of the function computed at $h$ that is represented by
a term $t(s)$ in $T(h),$
\item
$s$ is not a prime implicant of the function computed
  at either of the two direct predecessors $h_1$ and $h_2$ of $h,$
  and 
\item
  for the output gate $o$ such that $s$ is a prime
  implicant of the function computed at this gate,
  $t(s)$ is a subterm of the term $a(s)$
  in $T(o)$ assigned to $s$ (thus,
  there is a directed path connecting $h$  with $o$).
\end{enumerate}

Consider any AND gate $h$ of the circuit $C.$ Suppose that $|S_h|\ge 2.$
Then, $h$ jointly with its direct predecessors satisfy the assumptions
of Lemma \ref{lem: 2}. We obtain a contradiction with the $1$-multilinearity of $C.$
We conclude that $|S_h|\le 1.$
On the other hand, note
that for each $s\in PI(F),$ there must exist an AND gate $h$
in $C$ such that $s\in S_h.$
(To find such a gate $h$ start from the output gate $o$
computing the function in $F$ for which $s$ is a prime implicant
and $a(s)\in T(o),$ and iterate the following steps: check if the current gate
$g$ satisfies $s\in S_g,$ if not go to the direct
predecessor of $g$ that computes a function having
$s$ as a prime implicant represented by a subterm of $a(s)$.)
\qed
\end{proof}
\junk{
\begin{corollary}
A monotone arithmetic circuit computing
  a semi-disjoint bilinear arithmetic form with $p$ monomials
  has at least $p$ multiplications  gates.
\end{corollary}
\begin{proof}
  Consider a monotone arithmetic circuit $C$ computing the form.
  We may assume w.l.o.g. that e circuit does not have constant gates.
  Replace the multiplication gates with AND gates and the addition
  gates with OR gates, respectively. The resulting circuit is
  a monotone $1$-multilinear Boolean circuit computing a semi-disjoint
  Boolean form with $p$ prime implicants.
  \qed
\end{proof}

  If $t_1t_2$ is an
  implicant of the function computed at the gate $o$ then
  $t_1t_2$ has also to represent $u_1w_1$ so both $u_1$
  and $w_1$ have two occur at least twice in $t_1t(u_1w_1)t_2.$
  so the circuit $C$ cannot be $1$-multilinear.
  It follows that $t_1t_2$ represents either $u_1$ or $w_1$
  or $t_1t_2$ represents Boolean $1$, or just the terms
  do not exist so they can be represented by empty words. 
  The two latter cases cannot happen
  since the term $t_1t(u_1w_2)t_2$, where $t(u_1w_2)$
  is the term representing $u_1w_2$ in $T(h),$ is in $T(o)$
  but it cannot represent an implicant of the function computed
  at the gate $o,$  a contradiction. Also, when $t_1t_3$ represents $u_1$
  the term $t_1t(u_1w_2)t_2$ cannot represent an implicant
  of the function computed at $o.$ If $t_1t_2$ represents $w_1$
  then we consider the term $t_1t(u_2w_1)t-2$ in $T(o)$
  instead in order to get a contradiction.
  \end{proof}}

Our second result/observation in this subsection relies on Corollary 1
in \cite{L17}.

\begin{fact}\label{fact: centralmain}(Corollary 1 in \cite{L17})
Let $C$ be a monotone Boolean circuit 
of conjunction depth at most $d$ computing
a semi-disjoint bilinear form $F$ with
$p$ prime implicants.
The circuit $C$ has at least $\frac {p}{2^{2d}}$
AND gates.
\end{fact}

The following theorem is immediately implied by Theorem
\ref{fact: depth} and Fact \ref{fact: centralmain}.

\begin{theorem}\label{theo: qfirst}
  Let $C$ be a monotone $q$-multilinear Boolean circuit computing a
  semi-disjoint bilinear form $F$ with $p$ prime implicants.  The
  circuit $C$ has at least $\frac {p}{2^{4q-2}}$
  gates.
\end{theorem}
\junk{
In particular, Theorem \ref{theo: qfirst} yields the lower bound of
$\frac {n^2}{2^{4q-2}}$ on the number of gates in a monotone
$q$-multilinear Boolean circuit computing the $n$-dimensional Boolean
vector convolution.  The latter bound subsumes the best known lower
bound of $\Omega (n^2/\log ^6 n)$ on the size of monotone Boolean
circuits for this problem due to Grinchuk and Sergeev \cite{GS11}
as long as $q=o(\log \log n).$}

We can substantially subsume the lower bound of Theorem \ref{theo: qfirst}
  for larger $q$ by adhering to another lower bound on the size of
  restricted  monotone Boolean circuits computing semi-disjoint bilinear Boolean forms
   \cite{L17}.

We shall a call a class $K$ of monotone Boolean circuits
{\em k-nice} if (i) for each circuit $U\in K,$
for each output gate $o$ in $U$, each
non-zero term in $T(o)$ contains at most $k$ variables,
and (ii) $K$ is closed under the replacement of a gate in $U$
by a Boolean constant.

\begin{fact}\label{fact: warm}\cite{L17}.
Let $C$ be a 
monotone circuit that computes a semi-disjoint bilinear form $F$
on the variables $x_0,...,x_{n-1}$ and $y_0,...,y_{n-1},$
having $p$ prime implicants in total.
Suppose that $C$ belongs to a $k$-nice class $K$
and achieves a minimum size
among monotone circuits in $K$ that compute $F$. 
$C$ has at least $p/k^2$ AND gates.
\end{fact}

By combining Theorem 1 with Fact \ref{fact: warm},
we obtain the following lower bound that subsumes that of
Theorem \ref{theo: qfirst} asymptotically.
It is our second main result.

\begin{theorem}\label{theo: qsecond}
  Let $C$ be a monotone $q$-multilinear Boolean circuit computing a
  semi-disjoint bilinear form $F$ with $p$ prime implicants.  The
  circuit $C$ has $\Omega (\frac {p}{q^4})$
  gates.
\end{theorem}
\begin{proof}
  By Theorem 1, $C$ can be transformed into
  a monotone $q$-multilinear Boolean circuit $C'$
  computing the same semi-disjoint bilinear form
  such that all output terms of $C'$ include
  at most $2q$ variable occurrences and the size
  of $C'$ is at most $O(q^2)$ times larger than that
  of $C.$ . The set
  of monotone $q$-multilinear Boolean circuits whose
  output terms include at most $2q$ variable occurrences
  forms a $2q$-nice class. Hence, by Fact \ref{fact: warm},
  $C'$ has
  at least $\frac p {4q^2}$ gates.
    Consequently, $C$ has $\Omega(\frac p {q^4})$ gates.
    \qed
\end{proof}

In particular, Theorem \ref{theo: qsecond} yields the lower bound of
$\Omega(\frac {n^2}{q^4})$ on the size of a monotone
$q$-multilinear Boolean circuit computing the $n$-dimensional Boolean
vector convolution.  The latter bound subsumes the best known lower
bound of $\Omega (n^2/\log ^6 n)$ on the size of monotone Boolean
circuits for this problem due to Grinchuk and Sergeev \cite{GS11}
as long as $q=o(\log^{3/2} n).$

\subsection{Bounds for $Isol_{k,n}$}

We define the monotone Boolean function $Isol_{k,n}$
as follows.

Let $X=(x_{i,j,...,r})$ be an $k$-dimensional Boolean matrix such that
the indices $i,j,...,r$ are in $[n],$ where $[s]$ stands for
the set of positive natural numbers not exceeding $s.$ A {\em line} in $X$ is
any sequence of $n$ variables in $X$, where $k-1$ indices are
fixed and the index on the remaining position varies from
$1$ to $n.$ E.g., in the four-dimensional case, it can be
$x_{7,1,5,2},x_{7,2,5,2},...,x_{7,n,5,2}.$
\par
\noindent
$Isol_{k,n}(X)=1$ if and only if in each line in $X$ there is at least one $1.$
 For convenience, we say that a variable
 {\em guards} a line if (the matrix entry represented by) the variable
 belongs to the line.
 
We shall denote the set of shortest (prime) implicants of a Boolean
form $F$ by $SPI(F).$ Note that $SPI(Isol_{k,n})$ consists of (prime)
implicants of $Isol_{k,n}$ composed of $n^{k-1}$ distinct variables
guarding (i.e., belonging to) pairwise disjoint sets of $k$ lines in the
$k$-dimensional matrix $(x_{i_1,...,i_k})$.  In case $k=2,$ the
aforementioned implicants correspond to perfect matchings in
$K_{n,n}.$

\begin{lemma}
  The equality $|SPI(Isol_{k,n})|=
  \frac 1 {(n^{k-1})!} \prod_{i=0}^{n^{k-1}-1} (n^k-ikn+i(k-1))$ holds.
  \end{lemma}
  \begin{proof} \junk{It is easy to see by generalizing
    the relationship for $k=2$ that $|SPI(Isol_{k,n})$
    is equal to the number of perfect matchings in a $k$-uniform
    $k$-partite complete hypergraph, where each  part consists
    of $n$ vertices. The hyperedges in the hypergraph are just
    $k$-vertex subsets containing a vertex from each part.
    The number of aforementioned matchings is just the number
    of ways in which one can assign unique partners from
    the parts $2,3,...,k$ to the vertices in the first part.
    Thus, it is the product of the number of possible orderings
    of the second, third,..., $k$-th part, i.e., $(n!)^{k-1}.$}
    We can pick the first variable in a (prime) implicant in $SPI(Isol_{k,n})$
    in $n^k$ ways, the second one in $n^k-kn+(k-1)$ ways, since we have
    to avoid the lines already guarded by the first variable,
    similarly the third one in $n^k-2kn+2(k-1)$ ways, and so on.
    Finally, we need to divide the product of all the numbers of ways
    by $(n^{k-1})!$ as the order of the $n^{k-1}$ variables does not matter.
    \qed
  \end{proof}

Jukna showed that $Isol_{2,n}$ admits a monotone $2$-multilinear
Boolean circuit with $\le 2n^2$ gates \cite{SJ22}.  This upper bound
can be easily generalized to include an arbitrary $k\ge 2$.
The straightforward idea is to let the circuit to compute
for each line the disjunction of variables representing matrix entries on the
line and then to compute the conjunction of the disjunctions.
As each entry belongs to $k$ lines, each variable occurs $k$ times
in some terms representing prime implicants of $Isol_{k,n}$
produced by the circuit and on the other hand it never occurs more
than $k$ times in any term produced by the circuit.
Hence, we obtain the following remark.

\begin{remark}\label{fact: kmulti0}
  $Isol_{k,n}$ admits a $\Pi_2$ monotone strictly $k$-multilinear Boolean circuit
  with $kn^{k-1}(n-1)$ OR gates and $kn^{k-1}-1$ AND gates.
\end{remark}

  A higher alternation depth yields a possibility of lowering the number
of occurrences of at least a $\frac 1 n$ fraction of the input variables
to $1$ in the terms produced by a monotone Boolean circuit for $Isol_{k,n},$
without increasing the size of the circuit.

\begin{theorem}\label{theo: frac}
  $Isol_{k,n}$ admits a $\Pi_4$ monotone strictly $k$-multilinear Boolean circuit
  with $kn^{k-1}(n-2)+n^{k-1}$ OR gates and $kn^{k-1}-1$ AND gates,
  where a $\frac 1 n$ fraction of variables occur at most once in any
  term produced by the circuit.
\end{theorem}
\begin{proof}
Consider the $\Pi_2$ monotone $k$-multilinear circuit for $Isol_{k,n}$
given in the proof of Remark  \ref{fact: kmulti0}.  Pick a shortest prime implicant in
$SPI(Isol_{k,n})$ and let $Y$ be the set of variables forming it.
For each $y\in
Y,$ let $L_1(y),...,L_k(y)$ be the $k$ lines in the input matrix, where the
matrix entry represented by $y$ occurs.  Recall that for any distinct
$y,\ y',\in Y$ and any $1\le i, j \le k,$ $L_i(y)\neq L_j(y'),$ and
that for each line $L$ in the input matrix there is $y\in Y$ and $1\le
i \le k$ such that $L=L_i(y).$ In the conjunction of disjunctions
computed by the $\Pi_2$ circuit for $Isol_{k,n}$, for $y\in Y, 1\le i
\le k,$ there must be disjunctions $y \vee Y_i(y)$,
representing $L_i(y)$, for $1\le i \le k,$ respectively, where $Y_i(y)$ are disjunctions
of variables representing the remaining entries in the lines $L_i(y)$.
Thus, the $\Pi_2$ circuit computes $\bigwedge_{y\in
  Y}(\bigwedge_{1\le i \le k} y \vee Y_i(y))$.  This is equivalent to
$\bigwedge_{y\in Y}(y \vee \bigwedge_{1\le i \le k} Y_i(y))$.  So, we
can modify the $\Pi_2$ monotone Boolean circuit for $Isol_{k,n}$ to a
$\Pi_4$ one computing the latter conjunction. It is easy to see that
the modified circuit uses even a slightly smaller number of OR gates
and the same number of AND gates as the original one. It is still not
$(k-1)$-multilinear but at least a $\frac 1 n$ fraction of the input
variables (i.e., those in $Y$) occur at most once in terms
produced by the circuit.
\qed
\end{proof}

 As we have mentioned in the introduction,
Jukna observed an exponential gap between the size of
monotone $2$-multilinear Boolean circuits
and the size of monotone multilinear (i.e., $1$-multilinear) circuits
for $Isol_{2,n}$ \cite{SJ22}.
In the following, we show that a similar gap  holds
between $\Sigma_3$ monotone $k$-multilinear Boolean circuits
and $\Sigma_3$ monotone $(k-1)$-multilinear Boolean
circuits for $Isol_{k,n}$. Our result substantially subsumes
a corresponding result stated in Theorem 6 in \cite{L23} relying on additional assumptions
on the number of terms produced by the circuit.
We make use of the following observation.

\begin{lemma} \label{fact: common}.
  Consider a gate computing a disjunction of variables in a
  monotone Boolean circuit for $Isol_{k,n}$.  If the variables in the
  disjunction represent entries that do not belong to a common line in the input matrix
  then the gate can be replaced by the Boolean constant $1.$
\end{lemma}
\begin{proof}
  Suppose that there are two variables in the disjunction that do not
  share a line. Then, no variable occurrence from the disjunction is
  necessary to guard uniquely a line (i.e.,
  be the only variable belonging to the line) in any output term depending
  on the disjunction in order to make the term an implicant of
  $Isol_{k,n}.$ Otherwise, the sibling output term resulting from
  replacing the variable by another one belonging to the disjunction
  but not lying on the line would not be an implicant of $Isol_{k,n}$.
  Hence, the gate can be replaced by the Boolean constant $1.$
  Consequently, each pair of variables in the disjunction shares a
  line which implies that (the entries represented by) all variables in the disjunction occur on
  the same line in the matrix.
\qed
  \end{proof}

Our third main result follows.

\begin{theorem}\label{theo: last}
 Any $\Sigma_3$ monotone Boolean circuit for
 $Isol_{k,n}$
 has at least \\
 $|SPI(Isol_{k,n})|$ AND gates
or it is not $(k-1)$-multilinear.
\end{theorem}
\begin{proof}
We may assume w.l.o.g. that no gate in the circuit specified in the
  theorem statement can be replaced by a Boolean constant.  Suppose
  also that the circuit is $(k-1)$-multilinear.  We shall call a term
  produced at a gate of the circuit $q$-multilinear if no variable in
  the term occurs more than $q$ times.

  Consider an AND gate $g$ in the circuit whose
  $(k-1)$-multilinear terms represent the
  largest number of members in $SIP(Isol_{k,n})$ among all AND gates.
  We may assume w.l.o.g. that $g$ is a direct predecessor
  of some OR gate on the top level and there is
  a direct path from $g$ to the output gate in the circuit.

 The gate $g$ computes a conjunction of disjunctions of variables.  If
 the conjunction contains the number of distinct single variable
 disjunctions equal to the length of a shortest prime implicant of
 $Isol_{k,n}$ then the terms produced at the gate $g$ (i.e., belonging to $T(g)$) can represent
 only one such an implicant.
 Consequently, the number of AND gates has to be not less than
 $|SIP(Isol_{k,n})|$ in this case.

 Otherwise, some of the aforementioned disjunctions have to contain at
 least two variables.  By our assumptions and Lemma \ref{fact:
   common}, each of the at least two-variable disjunctions is composed
 of variables (representing entries) guarding (i.e., lying on) the
 same line. On the the hand, for each line at least one of the
 aforementioned disjunctions has to contain variables lying on this
 line.

 Consider $s\in SIP(Isol_{k,n})$ represented by a $(k-1)$--multilinear
 term produced at $g.$ Recall that each variable in $s$ guards a
 disjoint set of $k$ lines.  It follows that there is at least one set
 of $k$ lines that is uniquely guarded
 in $s$ by the same variable belonging solely to
 $\ell$ distinct at least two variable disjunctions. If $\ell < k$
 then if one picked a single alternative variable from each of the
 latter disjunctions, the picked variables could guard at most $k-1$
 of these lines.  Thus, the gate $g$ and consequently an output
 gate of the circuit would produce a term not representing an
 implicant of $Isol_{k,n}.$

It follows that the aforementioned variable occurs at lest $k$ times in the term
representing $s.$ We obtain a contradiction with the assumption on
$(k-1)$-multilinearity of the term and the circuit in this case. 
\qed
\end{proof}

\section{Final remark}
An ultimate goal would be to establish an exponential separation result similar to
that of Theorem \ref{theo: last} without the assumption on bounded
alternation depth of the circuit.  Unfortunately, the lower envelope
argument from \cite{JS82} seems to be helpful in separating only
monotone $2$-multilinear circuits from monotone $1$-multilinear ones
\cite{SJ22}.
\section*{Acknowledgments}
The first author thanks Susanna de Rezende for bringing attention to
the monotone Boolean circuit complexity of the Boolean permanent
problem studied in \cite{KV04,razb-perm} and valuable discussion on
monotone multilinear Boolean circuits.  Thanks also go to Stasys Jukna
for his valuable comments on monotone multilinear Boolean circuits.
The authors are also grateful to anonymous referees for their valuable comments on prior versions
of this paper.  The research was partially supported by Swedish
Research Council grant 621-2017-03750.

\vfill
\end{document}